\documentclass[smallextended]{svjour3}       
\smartqed 
\usepackage{graphicx}
\usepackage{enumitem}
\usepackage{lscape}
\usepackage{tikz}
\usepackage{tikz-qtree}
\usetikzlibrary{calc,trees,positioning,arrows,chains}
\usepackage{epstopdf}
\begin{document}

\title{Seeing The Whole Patient: Using Multi-Label Medical Text Classification Techniques to Enhance Predictions of Medical Codes }

\titlerunning{Multi-label Medical Text Classification}      

\author{Vithya~Yogarajan\and Jacob~Montiel \and Tony~Smith  \and Bernhard~Pfahringer}

\authorrunning{Yogarajan et al.} 

\institute{Vithya Yogarajan, Jacob Montiel, Tony Smith, Bernhard Pfahringer  \at
          Department of Computer Science, The University of Waikato \\
          \email{vy1@students.waikato.ac.nz} \\
}

\date{Received: date / Accepted: date}

\maketitle

\begin{abstract}
Machine learning-based multi-label medical text classifications can be used to enhance the understanding of the human body and aid the need for patient care. We present a broad study on clinical natural language processing techniques to maximise a feature representing text when predicting medical codes on patients with multi-morbidity. We present results of multi-label medical text classification problems with 18, 50 and 155 labels. We compare several variations to embeddings, text tagging, and pre-processing. For imbalanced data we show that labels which occur infrequently, benefit the most from additional features incorporated in embeddings. We also show that high dimensional embeddings pre-trained using health-related data present a significant improvement in a multi-label setting, similarly to the way they improve performance for binary classification. High dimensional embeddings from this research are made available for public use.\footnote{https://github.com/vithyayogarajan/medicaltext\_embeddings}          
\end{abstract}

\newpage
\section{Introduction} \label{intro}

The human body is a very complex system, and often patients admitted to hospitals with one initial prognosis or diagnosis have multiple related or unrelated chronic diseases,  referred to as multi-morbidity. Modern medical practice emphasises the need to understand the patient as a whole, as multi-morbidity increases the patient's overall burden of disease, and worsens prognosis \cite{flegel2018we,ryan2018beyond,hausmann2019sensitivity,aubert2019patterns,mori2019associations}. Multi-morbidity makes the diagnosis of each disease more complicated, and physicians may be less accurate in their diagnoses \cite{hausmann2019sensitivity}. The effects of different conditions may interact with each other, and complicate the management of each disease \cite{flegel2018we}. This, in turn, leads to poorer outcomes, such as increased preventable hospital re-admissions, overall hospital re-admissions, and increased total medical and long term care costs \cite{mori2019associations,aubert2019patterns}. For example, a patient newly diagnosed with HER2 (human epidermal growth factor receptor 2) positive breast cancer may also have underlying, possibly undiagnosed, heart failure. This can be crucial, as some treatments for breast cancer can cause cardiac damage. Accurately identifying the symptoms of heart failure allows the physician to best balance the risks and benefits of such treatments.  

Machine learning techniques have proven to aid medical advancements and enhance overall patient care. This research uses multi-label medical text classification techniques to improve prediction of the medical codes of patients with multi-morbidity. In single-label classification only one target variable is predicted per instance, i.e., each instance is assigned a class label out of 2 (binary) or more (multi-class) candidates. Whereas, in multi-label classification, the goal is to predict multiple output variables for each input instance.  In the above example, the patient is an instance with potential labels such as cancer, hypertension, heart failure, cholesterol and many more related and unrelated health complications. This research focuses on medical codes due to the availability of labels in the dataset. Medical codes such as international classification of diseases (ICD) are used as a way of classifying diseases, symptoms, signs and causes of diseases. Almost every health condition can be assigned a unique code. 

The focus of this research is to make use of free-form medical text. Free-form medical text such as discharge summaries, consultation notes and nurses notes are generally longitudinal and are rich sources of information about a patient's well-being and medical history. However, electronic health records (EHR) in free-form medical text present added complexity due to the nature of the content. EHRs in the free-form text contain an abundance of personal health identifiers which have to be carefully de-identified to avoid any ethical or legal issues \cite{yogarajan2020review}. Also, EHRs contain a large number of abbreviations and acronyms, which can be easily misinterpreted. For example, ``Mg'' is used to refer to magnesium, ``MG'' refers to Myestina gravis and ``mg'' refers to milligram.    

This research restricts itself to techniques that enable maximising the feature extraction of the medical text of the embedding layer. Embedding layer is a mapping of discrete variables to continuous  vectors, where the dimensional space of the categorical variables is reduced. The embedding layer is considered a significant component for text representation~\cite{goldberg2017neural}. Embeddings allow words to transform from isolated distinct symbols to mathematical representations, where the distance between vectors and distance between words can be equated, and behaviour between words can be generalised. We focus only on multi-label machine learning techniques commonly used in health-related information extraction tasks to better enhance the accuracy of predicting medical codes on patients with multi-morbidity.  

This paper extends the work on binary classification of medical codes presented in Yogarajan et al. (2020) \cite{yogarajan2020}. More specifically, in this paper:
\begin{itemize}
    \item [--] we acknowledge the multi-morbidity nature of patients, and we make use of the multi-label variations of medical text classification to enhance prediction of concurrent medical codes.
    \item [--] we present new embeddings on the health-related text and compare several variations to embeddings models when dealing with an imbalanced multi-label medical text classification problem.  
    \item [--] we analyse pre-processing of free-form medical text, given the nature of the medical text, and show that there are very minimum improvements to F-measure when medical text is pre-processed to that of the text `as is'.
    \item [--] we present a study exploring variations to tagging words including the traditional part-of-speech (POS). 
    \item [--] we provide a comparison of popular machine learning classifiers used in medical text classification.
    \item [--] we present a detailed study and discussion of results extended by varying the formations of embeddings, size of the embeddings and number of labels considered (18, 50 and 155) for the prediction of medical codes. 
    \item [--] we show that variations in embeddings, especially the dimensional size, influences the F-measure of the infrequent labels.
\end{itemize}  

The rest of the paper is structured as follows. Section 2 presents related work. This is followed by a brief overview of medical codes in Section 3. Details of the machine learning techniques and experimental methodology are provided in part 4. Section 4 also presents an overview of the data used for experiments. This is followed by results, where a detailed subsection of results are given for 18 label case, followed by 50 and 155 labels. The paper is concluded with discussions and suggestions for future work.

\section{Related Work} \label{sec:lit}

Developments in machine learning, especially deep learning, have influenced the advancements in many fields, including health applications. The rapid growth in computational power and the availability of EHR are the main reasons for such changes. Rule-based systems have been the most favoured option by health professionals, with systems such as cTAKES and MetaMap considered the leading information extraction tools \cite{savova2010mayo,garla2011yale,liu2013integrated,reategui2018comparison,yang2017uevora}. However, recently there is a shift towards favouring machine learning, more specifically deep learning-based models. 

Table \ref{table:lit} presents examples of recent developments in predicting medical codes. All systems are based on variations of deep learning models. The number of ICD-9 codes, i.e. the number of labels, used varies across systems with the best reported F1 measures around the 0.4 to 0.6 range. The number of labels and the frequency of the chosen labels influence the F1 score, with top 50 ICD-9 codes generally leading to higher F-measure. MIMIC III (Medical Information Mart for Intensive Care III) is the biggest publicly accessible de-identified dataset and is the most popular free-form medical text used in many applications, including predicting medical codes   \cite{purushotham2017benchmark,johnson2017reproducibility,goldberger2000physiobank,data2016secondary} (also evident in Table \ref{table:lit}).    

\begin{table}[t]
\caption{Examples of the most recent systems for predicting ICD-9 codes are presented. Here CNN refers to convolutional neural network, LSTM to Long short-term memory, Bi-GRU to bidirectional Gated Recurrent Unit and DR-CAML to Description Regularized - Convolutional Attention for Multi-label classification. * Du et al. (2019)~\cite{du2019ml} do not specify best micro average or macro average F score. }
\label{table:lit}
   \begin{center}
\resizebox{\linewidth}{!}{
    \begin{tabular}{lllll}
    \hline\noalign{\smallskip}
System & Methods & Data & Best Score  & Details \\
\noalign{\smallskip}\hline\noalign{\smallskip}
 Zeng et al. (2019)~\cite{zeng2019automatic} & Deep transfer learning & MIMIC III & micro avg   & most frequent 200 labels\\
&Multi-scale CNN & & F1 =  0.420  & \\\noalign{\smallskip}
Du et al. (2019)~\cite{du2019ml} & ML-Net, ELMo based, & MIMIC III & Best* & \\
&  & & F1 = 0.428 & 70 labels \\
& LSTM & &  &  \\\noalign{\smallskip}
Baumel et al. (2018)~\cite{baumel2018multi} & Hierarchical Attention  & MIMIC III & micro avg & \\
&  Bi-GRU & & F1 = 0.405 & 6527 labels \\
& & & F1 = 0.559 & 1047 labels \\
\noalign{\smallskip}
Mullenbach et al. (2018)~\cite{mullenbach2018explainable} & CNN based, Word2Vec & MIMIC III & micro avg  & \\
&DR-CAML & & F1 = 0.633 & most frequent 50 labels \\
&CAML & & F1 = 0.539 & 8922 labels \\
& & & macro avg & \\ 
&CAML & &F1 = 0.088 & 8922 labels \\\noalign{\smallskip}
Li et al. (2018)~\cite{li2018automated} & DeepLabeler & MIMIC III & micro avg  & 6984 labels \\
& CNN, Doc2Vec & &F1 = 0.408 & \\ \noalign{\smallskip}
Rios and Kavuluru (2018)~\cite{rios2018emr} &CNN, few-shot learning & MIMIC III & micro avg & 6932 labels \\
&Skip-gram embeddings & & F1 = 0.468 & \\
 \noalign{\smallskip}\hline\noalign{\smallskip}
\end{tabular}}
\end{center}
\end{table}

Embeddings are the popular method used to represent text in a neural network, and all systems presented in Table \ref{table:lit} use embeddings from algorithms such as Word2Vec, Doc2Vec and ELMo to represent free-form medical text. Yogarajan et al. (2020) \cite{yogarajan2020} used fastText to obtain embeddings and presented comparisons with published embeddings, both for general text and health-related text trained models. Embeddings trained on health-related text perform better than those trained on general text, and higher dimensions perform better when top-level ICD-9 groups are considered as an individual binary problem \cite{yogarajan2020}. Huggard et al. (2019) \cite{huggard2019feature} also show that embeddings obtained from fastText result in significantly higher F-measure on the biomedical name entity recognition when compared to other embeddings such as that of ELMo.  

We restrict this research to enhancing embeddings in a multi-label prediction setting. Our findings in this research will aid the development of better performing neural networks. All systems presented in Table \ref{table:lit}, and other deep learning-based models, focus predominantly on the complexity of the deep learning algorithm and very little on the representation of the text and pre-processing of the text. Although we acknowledge the need for such developments, in this paper, we constrain ourselves to text representations as this is vital to improving predictive performance for health records. This is so to avoid using the same baseline recipe for the embeddings layer where the size of the embedding is the same; generally, 100 dimensions, and pre-processing steps are also the same \cite{zeng2019automatic,mullenbach2018explainable}.

\section{Medical Codes} \label{sec:medcodes}

ICD codes are widely used to describe diagnoses of patients, and are used to classify diseases, symptoms, and also causes of diseases~\cite{jensen2012}. Many countries use ICD codes for billing purposes, as does the USA where insurance must cover the cost of patient care. ICD codes also provide insights on multi-morbidity of patients. We focus on predicting ICD-9 codes in this paper due to the availability of labels in the data. Generally, hospitals manually assign the correct codes to patient records based on doctors' clinical diagnosis notes. This requires expert knowledge and is time-consuming. Hence, the use of advancements in machine learning to predict ICD codes from free-form medical text has become an important research avenue.

\begin{figure}[t]
    \centering
\resizebox{\linewidth}{!}{
\begin{tikzpicture}[level distance=20mm,sibling distance=26mm]
\tikzset{edge from parent/.style= 
            {thick, draw},
         every tree node/.style=
            {draw,minimum width=0.5in,text width=0.5in,align=center}}
\node[draw] at (-11,-1.75) {\large{18 top-level groups}};
\node[draw] at (-11,-3.75) {\large{167 sub-level groups}};
\node {\large{ICD-9}}
    child {node {\large{001-139 (inf)}}
        child {node {\large{001-009 (inf1)}}
            child{node {\large{001}}
                child{node {\large{001.0}}}
                child{node {\large{001.1}}}
                child{node {\large{001.9}}}
            }
            child{node {\large{...}}}
            child{node {\large{009}}}
        }
        child {node {\large{...}}}
        child {node {\large{137-139 (inf16)}}} 
    }
    child {node {\large{140-239 (neop)}}}
    child {node {\large{...}}}
    child {node {\large{E \& V}}
    };
\end{tikzpicture}}
\caption{ICD-9 Hierarchy. The first split contains 18 groups (which we refer to as the top level ICD-9 grouping). These top level ICD-9 groups then splits into 167 sub-groups. Leaves represents the individual ICD-9 codes. All of the top level groups split into individual ICD-9 codes. E \& V refers to external causes of injury and supplemental classification and the ICD-9 codes that belongs to this group contains codes starting with E or V.}
\label{fig:icd9hir}
\end{figure}
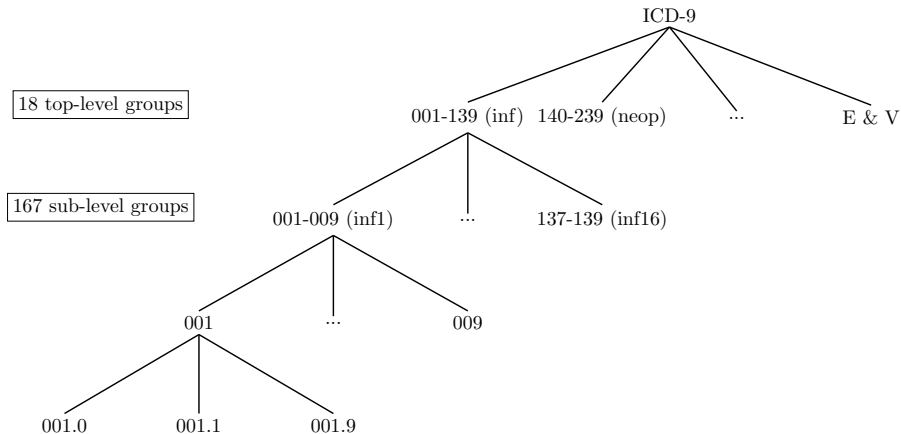

There are roughly 13,000 ICD-9 codes and their definitions follow a hierarchical structure. Figure \ref{fig:icd9hir} presents the tree structure of ICD-9. At the top level, ICD-9 codes can be grouped into 18 main categories, which then divide into 167 sub-groups and finishes with roughly 13,000 individual codes.   

\section{Experimental Methodology} \label{sec:methods}

This section presents an overview of the data used in experiments and for  training embeddings. We provide an overview of the machine learning and natural language processing techniques. The  details of the embeddings used in this research are also presented.  

\subsection{Data}

This research makes use of the medical text data of more than 50,000 patients presented in the publicly available medical database Medical Information Mart for Intensive Care (MIMIC-III) \cite{johnson2016mimic,goldberger2000physiobank,data2016secondary}. MIMIC III contains de-identified medical free-form text among other forms of medical data of patients admitted in critical care units at the Beth Israel Deaconess Medical Center between 2001 and 2012. MIMIC III contains 15 categories of notes in the free-form text, including discharge summaries, nursing notes, nutrition notes and social work notes. More than 90\% of the unique hospital admissions contain at least one discharge summary, with many including more than one. We make use of discharge summaries of individual hospital admissions in this research.

There are 6,984 distinct diagnosis ICD-9 codes and 2,032 distinct procedure ICD-9 codes reported in MIMIC III, among more than 50,000 patient admission records found in this database. Patient records in MIMIC III typically have more than one code assigned. However, the frequency of ICD-9 codes is extremely unevenly spread, with a large proportion of the ICD-9 codes occurring infrequently. Table \ref{table:summary_stats} provides an overview of the frequency of ICD-9 codes and ICD-9 groupings. We focus on the top-level and sub-level groups in this study, along with the first 50 highest frequently occurring individual ICD-9 codes. Table \ref{table:summary_stats} also presents frequency ranks of ICD-9 codes and sub-groups to showcase the unbalanced nature of the data. This biased nature is primarily because MIMIC III data were obtained from patients admitted in critical care.

\begin{table}[t]
\caption{Percentage of occurrence of ICD-9 codes and ICD-9 groupings (top-level and sub-level) in MIMIC III discharge summaries of unique hospital admissions is presented. For top level all 18 groups are presented. For sub-level group and top 50 ICD-9 codes only selected frequencies are presented with corresponding ranking of these frequencies (ordered highest to lowest). The total number of hospital admissions with a recorded discharge summary in MIMIC III is 52,710. Only ICD-9 codes or groups that occurred in $>=$ 10 unique hospital admissions are included.  }
\label{table:summary_stats}
   \begin{center}
\resizebox{\linewidth}{!}{
    \begin{tabular}{lllllr}
    \hline\noalign{\smallskip}
\multicolumn{6}{c}{ICD-9 Top Level Grouping: 18 Groups} \\\noalign{\smallskip}
Group & \% & Group & \% & Group & \% \\
\noalign{\smallskip}\hline\noalign{\smallskip}
circ (390-459)  & 78.40 & diges (520-579) & 38.80 & musc (710-739) & 17.99 \\
e+v (E- \& V-)   & 69.09 & bld (280-289) & 33.56 & pren (760-779) & 17.07 \\
endo (240-279)  & 66.51 & symp (780-799) & 31.36 & neop (140-239) & 16.37 \\
resp (460-519)  & 46.63 & ment (290-319) & 29.66 & skin (680-709) & 12.02 \\
inj (800-999) & 41.42 & nerv (320-389) & 29.10 & cong (740-759) & 5.41 \\
gen (580-629) & 40.29 & inf (001-139) & 26.96 & preg (630-679)  & 0.31 \\
 \noalign{\smallskip}\hline\noalign{\smallskip}
 \multicolumn{6}{c}{Top 50 ICD-9 codes} \\ \noalign{\smallskip}
{Frequency} & ICD-9 code  & \% & {Frequency} & ICD-9 code  & \% \\
rank & & & rank & & \\
 \noalign{\smallskip}\hline\noalign{\smallskip}
1 & 401.9 &35.13 & 20 & V05.3& 9.81\\
5 &414.01 &21.09 & 30 & 496.0&7.52 \\
10 &967.1 &15.11 & 40 &305.1 & 5.70\\
15 &599.0 &11.13 & 50 & V15.82 & 4.77\\
\noalign{\smallskip}\hline\noalign{\smallskip}
\multicolumn{6}{c}{ICD-9 Sub-Level Grouping: 155 Groups} \\ \noalign{\smallskip}
{Frequency} & Sub-group  & \% & {Frequency} & Sub-group  & \% \\
rank & & & rank & & \\
\noalign{\smallskip}\hline\noalign{\smallskip}
1 & endo4 (270-279)  & 52.13 & 50 & blood3 (288-289) & 4.28 \\ 
5 & symp1 (780-789) & 30.86 & 75 & inj4 (820-829) & 1.81 \\
10 & v6 (V40-V49)  & 25.76 & 100 & cong7 (753-753) & 0.56 \\
25 &diges6 (560-569)  & 11.47 & 155 & v12 (V86-V86) & 0.02\\
 \noalign{\smallskip}\hline\noalign{\smallskip}
\end{tabular}}
\end{center}
\end{table}

\subsection{Training Embeddings}\label{sec:embeddings}

Representing words as embeddings is a common mechanism used in language processing \cite{goldberg2017neural}. Embeddings obtained from algorithms such as Word2Vec, Glove and fastText are used in text classification tasks, including medical applications. This research makes use of fastText \cite{bojanowski2016enriching,joulin2016bag,joulin2016fasttext} where words are represented as a bag of character $n$-grams, and word embeddings are obtained by summing these representations. This feature gives fastText the ability to produce vectors for words that are misspelt or concatenated. The nature of free-form medical text does benefit from this feature of fastText embeddings, and there are examples of medical applications where embeddings from fastText are shown to outperform other similar algorithms \cite{huggard2019feature}.

Medical codes for patients admitted to the hospital are labelled at individual admission or patient level documents rather than single words in a health record. Hence, we predict medical codes for entire documents, in this case, discharge summaries of unique hospital admission. For this research, document embeddings are obtained by computing the vector sum of the embeddings for each word in the document. This vector sum is then normalised to have length one, ensuring documents with different lengths have representations of similar magnitudes.

For comparison, our embeddings are trained to the exact same specifications as the fastText embeddings W300 presented in Grave et. al. (2018)~\cite{grave2018learning}. Table \ref{tab:models} presents details of the embedding used in this research with details of dimensional size, source data, model size and training time\footnote{Processing was run on a 4 core Intel i7-6700K CPU @ 4.00GHz with 64GB of RAM.}. The word embeddings are trained using CBOW (T300, T600) and Skip-gram (T300SG, T600SG), with character n-grams of length 5, a window of size 5 and ten negative samples per positive sample. The learning rate used for training these models is 0.05.

We make use of the data provided by TREC 2017 competitions~\cite{roberts2017overview} to train our embeddings. TREC 2017 provides an extensive 24G of health-related data. TREC data contains 26.8 million published abstracts of medical literature listed on PubMed Central, 241,006 clinical trials documents, and 70,025 abstracts from recent proceedings focused on cancer therapy from the American Association for Cancer Research and the American Society of Clinical Oncology~\cite{roberts2017overview}. 

\begin{table}[t]
\caption{Word embeddings used in this research are presented, with dimension details, training times and embeddings model sizes.  }
\label{tab:models}
\begin{center}
\resizebox{\linewidth}{!}{
\begin{tabular}{lrlrr}
\hline\noalign{\smallskip}
Models & Dimensions & Source Data& Train Time & Model Size \\
\noalign{\smallskip}\hline\noalign{\smallskip}
W300~\cite{grave2018learning} & 300 & Wiki  & - & 7G \\
T300~\cite{yogarajan2020} & 300 & TREC & 7 hours & 13G \\
T300SG & 300 & TREC & 28 hours & 13G \\
T600~\cite{yogarajan2020} & 600 & TREC & 13 hours & 23G \\
T600SG & 600 & TREC & 51 hours & 23G \\
\noalign{\smallskip}\hline
\end{tabular}}
\end{center}
\end{table}

\subsection{Multi-label Classifiers}

Generally, a given medical record is annotated with multiple tags for different diagnoses, procedures or treatments. That is, from a machine learning perspective, health text coding is a multi-label classification problem, where one text may belong to more than one label. For example, many ICD codes exist for matters that relate to hypertension or diabetes, and such illnesses often co-occur in individual patients, but they also occur independently. Thus, it would be useful to be able to classify a particular health text to one or the other, or both, or neither of these categories.  Moreover, with approximately 13,000 ICD-9 categories for diagnoses and treatments that can combine almost arbitrarily for individual patients, the problem of multi-label classification for health records can be extraordinarily large.

In this section, we provide an overview of multi-label classifiers used for the experiments in this paper. For more details on these methods, see \cite{read2016}. 

\subsubsection{Binary relevance (BR)}
The first and simplest multi-label classification algorithm used here is called binary relevance (BR)~\cite{godbole2004discriminative,tsoumakas2007multi}. A separate binary classification model is created for each label, such that any text with that label is a positive instance, negative otherwise (i.e. one versus all). To predict the labels for a new text, each classifier decides if the text is in or out the class it has been trained to recognise, and the overall output for the new text is the set of all positive labels. Note that binary relevance ignores any potential relationships between labels. 

\subsubsection{Classifier Chains (CC)}
BR models make their predictions independently. However, as seen with the earlier example of the strong correlation between diabetes and hypertension, a model could possibly benefit from the result of another when making its own decision.  Accordingly, BR models can be `chained' 
together into a sequence such that the predictions made by earlier classifiers are made available as additional features for the next classifier.  Such a configuration is unsurprisingly called a classifier chain (CC)~\cite{JesseRead2009,JesseRead2011}.

\subsubsection{Ensemble of classifier chains (ECC)}
The order of predictions in a classifier chain affects what advice later models have available from preceding ones when it's their turn to make a judgment.  This is a problem for multi-label classification in general, but particularly so for health records where dependencies between ICD codes are myriad, complex, and sometimes quite strong.  One way to mitigate the problem of choosing a poor ordering is to create a collection of classifier chains that are each ordered randomly, then make final predictions by polling the results of all chains.  Such a collection is called an ensemble of classifier chains (ECC)~\cite{JesseRead2009}.

\subsubsection{Multi-label k-nearest neighbor classifier (MLkNN)}
MLkNN~\cite{zhang2005k} is a multi-label variant of the standard k-Nearest Neighbor (kNN) algorithm, that predicts the set of the most common labels among the $k$-nearest neighbours. To guard against any anomalies inside a neighbourhood, a Bayesian calibration step refines the raw predictions. An important characteristic of this approach is its excellent scalability with respect to the number of labels: the set of nearest neighbors needs to be calculated only once for a given query text.

\subsubsection{Neural Networks}

As emphasised earlier, the focus of this research is only at embeddings layers, and we use the most commonly used multi-label classifiers for prediction. However, the outcome of this research can be incorporated into a neural network, where embeddings layers are generally used to represent text \cite{goldberg2017neural}. Furthermore, in 2019 very recent NLP techniques like BERT (Bidirectional Encoder Representations from Transformers) and BioBERT, showed significant improvements on some other biomedical tasks \cite{lee2019ncuee,abacha2019overview}. These are all worthy avenues for future research.    

\subsubsection{MEKA and base classifiers}
All of the classification results presented in this research were carried out using MEKA~\cite{read2016}: an open-source Java system specifically designed to support multi-label classification experiments.  MEKA includes almost all widely-used algorithms and evaluation metrics.  The default algorithm for each class (i.e. the base classifier) within MEKA is logistic regression, and this is used for the majority of our experiments; however, stochastic gradient descent (SGD) is used for tests with ECC, with ensembles of 50, 100 and 500 randomly ordered classifier chains.

\subsection{Statistical assessment of differences}
We perform non-parametric tests to verify if there are statistically significant differences between algorithms, as described in~\cite{Demsar2006,Garcia2009}. First we use Davenport's corrected Friedman test with $\alpha=0.05$ to check if we can safely reject the null hypothesis that all algorithms perform the same. If there are differences, we proceed with the post-hoc Nemenyi test to determine the critical difference (CD) that serves to identify algorithms with different performance. We include the critical difference plots in our results.

\subsection{FastText Parameters} \label{sec:fastpar}

As mentioned in Section \ref{sec:embeddings} models used in this research are trained to the exact same specifications as general text trained published models. We also present a comparison of variations of specifications for training embeddings for a multi-label medical text classification problem. The combination of variations to parameter choices are presented in Table \ref{table:parameter}. The learning rate used to train all of the variations presented is 0.05. For simplicity all dimensions were set to 50. The two word representation models are Skip-gram and CBOW. It is important to note that these combinations result in 18 different embeddings models, and only a selected sub-set of the experimental results are presented in this paper. 

\begin{table}[h]
\caption{Variations of parameter choices for embeddings trained using fastText are presented. These options are used for both CBOW and Skip-gram. Option ``I'' contains the exact same parameters choices as the published model W300. ``neg'' refers to the number of negative samples per positive sample. [minn, maxn] refers to minimum length and maximum length respectively. *Option V sets maxn = 0, this means no sub-words will be used by fastText. Hence, the model should give similar results to that of word2vec.}
\label{table:parameter}
   \begin{center}
\resizebox{\linewidth}{!}{
    \begin{tabular}{lccccclc}
    \hline\noalign{\smallskip}
Option & Dimensions & Window  & neg & \multicolumn{2}{c}{Character $n$-gram} & Loss Function  & Epoch \\ 
& &Size & & minn & maxn & & \\ \noalign{\smallskip}\hline\noalign{\smallskip}   
I & 50 & 5 & 10 & 5 & 5 & softmax & 5 \\
II & 50 & 3 & 10 & 5 & 5 & softmax & 5 \\
III & 50 & 7 & 10 & 5 & 5 & softmax & 5 \\
IV & 50 &5 & 5& 5& 5& softmax & 5 \\
V* & 50 &5 &10 &0 & 0 & softmax & 5\\
VI & 50 &5 &10 &3 &3 & softmax & 5\\
VII & 50 &5 &10 &5 &5 & hierarchical softmax & 5 \\
VIII & 50 &5 &10 &5 &5 &negative sampling & 5 \\
IX & 50 & 5 & 10 & 5 & 5 & softmax & 10 \\
\hline\noalign{\smallskip}
\end{tabular}}
\end{center}
\end{table}

The Continuous Bag-of-Words Model (CBOW) \cite{DBLP:journals/corr/abs-1301-3781} is similar to a feed-forward neural network language model with non-linear hidden layer removed and the projection layer being shared for all words. CBOW predicts the current word from the surrounding words. The Skip-gram architecture \cite{DBLP:journals/corr/abs-1301-3781} is similar to that of CBOW, but Skip-gram uses the current word to predict the words before and after within a given range. For example, for a sentence ``Male patient is admitted to the hospital'', CBOW predicts the word ``admitted'' using the source context words (``Male'', ``patient'', ``is'', ``to'', ``the'', ``hospital''), whereas Skip-gram predicts context words like  ``patient'' or ``hospital'' for the source word ``admitted''.

\subsection{Pre-processing Text Data}\label{sec:preprocess}

We present a comparison of F-measures between pre-processed discharge summary and text `as is' as presented by MIMIC III data. MIMIC III data has been de-identified and pre-processed before being released for research access. Also, most models developed using MIMIC pre-process the text and truncate the maximum number of words \cite{mullenbach2018explainable}. Text pre-processing includes removal of tokens without alphabetic characters, down-casing all tokens, removal of punctuation and truncating the number of tokens in a given discharge summary. 

On the other hand, experiments presented in this research use MIMIC III discharge summaries `as is' with minimum pre-processing. This allows us to maximise the use of features in medical free-form text as embeddings are case sensitive. It also avoids the meanings of abbreviations and acronyms used in a medical context being altered. An example of text `as is' followed by an example of pre-processed text is presented below: 
\vspace{10pt}
\hrule
\begin{quote}
    Medicine HISTORY OF PRESENT ILLNESS: This is an 81-year-old female with a history of emphysema, presents with 3 days of shortness of breath thought by her primary care. Medications on Admission: Omeprazole 20 mg daily, Furosemide 10mg daily. Tablet Sustained Release 24 hr PO once a day.
\end{quote}
\hrule

\begin{quote}
    medicine history of present illness this is an 81yearold female with a history of emphysema presents with days of shortness of breath thought by her primary care
    medications on admission omeprazole mg daily furosemide 10mg daily tablet sustained release hr po once a day
\end{quote}
\hrule
\vspace{10pt}

\subsection{Concatenating Embeddings} \label{sec:concat}

We explore the option of splitting the free-form medical data into sections and concatenating the embeddings. The discharge summary is split into seven logical sections: Admission Date, Past Medical History, Pertinent Results, Brief Hospital Course, Medications on Admission, Discharge Diagnosis and Followup Instructions. Embeddings for each section can be obtained and concatenated. For example, if a 50 dimensional embeddings model is used the resulting concatenated embedding has 350 dimensions. If the discharge summary does not include any of the sub-sections mentioned above, then the respect embeddings are all zeros. For hospital admissions with more than one available discharge summary, all the summaries are first embedded independently, and then averaged into one final embedding.

Another variation considers concatenating statistical outcomes of the embeddings from each of the sections of a given hospital admission. For these experiments we look at the minimum, maximum, mean, standard deviation, lower quartile and upper quartile of the embeddings, and hence the resulting embedding will have six times as many dimensions as the original one.

\subsection{Tagging Words} \label{sec:tagging}

We explore two variations of tagging words in medical free-form text. Part-of-speech (POS) is a technique used where the syntactic categories of words in a given sentence are identified automatically. Common examples of such POS tags are: noun, verb, adjective, adverb, pronoun, preposition, conjunction and interjection. We make use of Natural Language Toolkit (NLTK\footnote{http://nltk.org/})~\cite{NLTK09} POS tagger, where if the input text is:
\begin{quote}
History of Present Illness 54 year old female with recent diagnosis of ulcerative colitis on mercaptopurine 
\end{quote}
\noindent output is:
\begin{quote}
HistoryNN ofIN PresentNNP IllnessNNP 54CD yearNN oldJJ femaleNN withIN recentJJ 
diagnosisNN ofIN ulcerativeJJ colitisNN onIN mercaptopurineJJ 
\end{quote}
where NN indicates a noun, IN is referring to preposition or conjunction, NNP is referring to a proper noun, CD is referring to numeral and JJ is referring to adjective or numeral. 

Also, we tag the words of MIMIC III discharge summaries using the text splits presented in Section \ref{sec:concat}. Tokens in each of these sections are tagged with 0\_, 1\_, 2\_, 3\_, 4\_, 5\_, 6\_ for text in the seven splits Admission Date, Past Medical History, Pertinent Results, Brief Hospital Course, Medications on Admission, Discharge Diagnosis and Followup Instructions respectively. 

\section{Results}

This section presents results for the top level ICD-9 grouping, the sub-level grouping, and the overall top 50 highest frequency ICD-9 codes where the number of labels are 18, 155, and 50 respectively. The top level ICD-9 groupings are primarily used to present comparisons and detailed results for the multi-label medical text classification techniques mentioned in Section \ref{sec:methods}. Results are primarily presented at the level of individual labels to enable better understanding of the imbalanced nature of the data and to observe improvements in F-measure. We also present  micro-averaged and macro-averaged F1 scores to facilitate comparisons across the variations in embeddings to the overall system.      

\subsection{Top-Level Groups of Medical Codes}\label{sec:top}

This section presents results for the 18 top-level ICD-9 groups, as mentioned in Table \ref{table:summary_stats}. We also present comparisons of 18 groups treated as individual binary problems, as presented in Yogarajan et al. (2020) \cite{yogarajan2020}, where appropriate. The results in this section are aligned with the experimental methodology as described above. 

\subsubsection{Comparing Multi-label Classifiers}

\begin{figure}[ht]
    \centering
    \includegraphics[width=.9\textwidth]{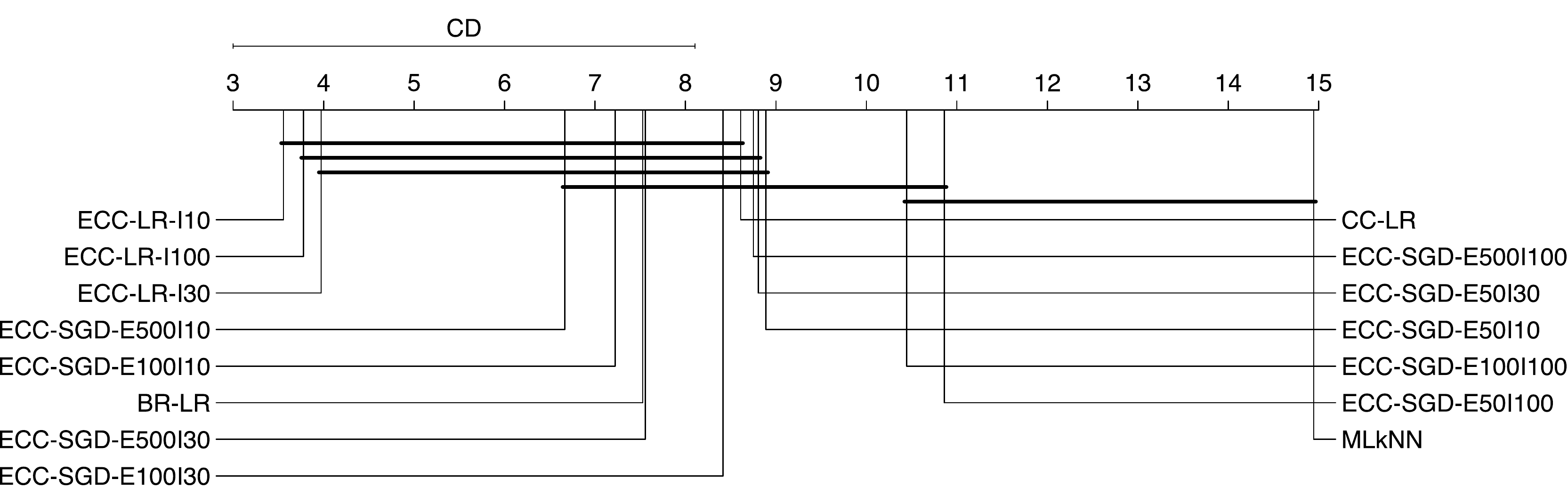}
    \caption{Critical difference plots. Nemenyi post-hoc test (95\% confidence level), identifying statistical differences between methods in our tests.}
    \label{fig:cd_plot_class_final}
\end{figure}

\begin{table}[t]
\caption{Comparison of F-measures for the 18 top level ICD-9 groups is presented for varying multi-label classifiers. I indicates number of iterations and E is the number of epochs to perform. T300 is used for embeddings. Bold is used to indicate F-measures better than that of the BR-LR, and underline to indicate the best F-measure across all presented.}
\resizebox{\linewidth}{!}{
\begin{tabular}{lccccllcl}
\hline\noalign{\smallskip}
ICD-9   &BR-LR & MLkNN & CC-LR  &  \multicolumn{3}{c}{ECC-SGD} &  \multicolumn{2}{c}{ECC-LR}   \\
groups  &  &  &  & \multicolumn{3}{c}{Best case}   & \multicolumn{2}{c}{Best case}   \\
&    &  &  & F1 & E = & I = & F1 & I = \\
\noalign{\smallskip}\hline\noalign{\smallskip}
circ & 0.932 & 0.921 & {0.932} & \underline{\textbf{0.933}} & 100 &  10  & {\textbf{0.932}} & 30, 100 \\ 
e+v    & 0.829 & 0.823 & 0.812 & \underline{\textbf{0.831}} & 100 & 30  & \textbf{0.830} & 30, 100\\ 
endo & 0.848 & 0.839 & 0.848  & \underline{\textbf{0.851}} & 50 & 100 & 0.848 &10, 30, 100 \\ 
resp   & 0.777 &0.703 & 0.770 & \underline{\textbf{0.784}} & 500 & 10 & \textbf{0.782} & 30 \\ 
inj  & 0.662 & 0.590 & 0.653 & \textbf{0.683} & 500 & 10& \underline{\textbf{0.686}} & 30 \\ 
gen  & 0.731 & 0.657 & 0.731 &\textbf{ 0.735} & 500 & 30 & \underline{\textbf{0.739}} & 10, 30, 100 \\ 
diges  & 0.696 & 0.600 & 0.694 & \textbf{0.706} & 50& 30& \underline{\textbf{0.713}} & 10 \\ 
bld   & {0.571} & 0.494 & {0.571}& \textbf{0.577} &50 &100 & \underline{\textbf{0.612}} & 10\\ 
symp  & 0.487 & 0.361 &0.463 & \textbf{0.489} &50 &10 & \underline{\textbf{0.552}} & 10\\ 
ment   & 0.542 & 0.299 & 0.538 & \textbf{0.562} &500 &30 & \underline{\textbf{0.590}} & 10\\ 
nerv   & 0.543  & 0.376 & 0.522 & 0.530 &100 & 10 & \underline{\textbf{0.589}} & 10\\ 
inf  & 0.647 & 0.547 & \textbf{0.651} & \textbf{0.667} &500 & 30& \underline{\textbf{0.683}} & 10 \\ 
musc  & {0.298} & 0.086 & \textbf{0.302}& 0.272 & 500&30 & \underline{\textbf{0.410}} & 10\\ 
pren   &  \underline{0.594} &0.575 & 0.592 &0.588 & 500&10 & 0.590 & 10\\ 
neop  &  0.703 & 0.500 & 0.705 & \textbf{0.709}  & 500&10 & \underline{\textbf{0.718}} &10 \\ 
skin    &  {0.347}  & 0.075 & \textbf{0.349} &0.328 &500 &10 & \underline{\textbf{0.413}} &10 \\ 
cong  & 0.384 & 0.294 & 0.383 & 0.361 &500 &100 & \underline{\textbf{0.449}} & 30\\ 
preg   &  \underline{{0.592}} & 0.267 & {0.572} &  0.542 &500 &100 & 0.514 & 100\\ \noalign{\smallskip}\hline
\end{tabular}}
    \label{tab:18_clas_final}
\end{table}
Table \ref{tab:18_clas_final} presents a comparison between several multi-label classifiers to predict the 18 top-level ICD-9 groups. Critical difference plots are available in Figure~\ref{fig:cd_plot_class_final}. Performance when considering the 18 groups as individual binary classification problems is also presented. All experiments use the T300 word embedding and 10-fold cross validation. As anticipated, using multi-label variations does provide advantage over the individual binary classification case. Evidently, for most ICD-9 groups ECC using with logistic regression (LR) performs best. Optimising the number of iterations and epochs can improve F-measure results. ECC-LR with a ridge value of $R = 1$ and the number of iterations $I = 10$ achieves the best results overall. Experiments across a range of different ridge values provided almost identical values for F-measure, hence only a ridge value of 1 is included in Table \ref{tab:18_clas_final}.

\subsubsection{FastText Parameter Choices} \label{sec:fastparresult}

\begin{table}[t]
    \centering
        \caption{Comparison of F-measures for top level ICD-9 groups for embeddings obtained by varying fastText parameters, as presented in Table \ref{table:parameter}. Options I to IX match that of Section \ref{sec:fastpar}. Embeddings are trained on TREC data and the classifier used for experiments is BR with logistic regression and $R=1$. Best F1 score for CBOW models are presented on the left. Bold is used to indicate F-measures better than the best CBOW F1 score, and underline is used to indicate the best F1 across the options presented.  }
    \resizebox{\linewidth}{!}{    
    \begin{tabular}{llccccccccc}\hline\noalign{\smallskip}
 ICD-9   & {CBOW} & \multicolumn{9}{c}{Skip-gram} \\
groups&Best F1&I&II&III&IV&V&VI&VII&VIII&IX\\\noalign{\smallskip}\hline\noalign{\smallskip}
circ&0.924 &\textbf{0.925}&0.923&\underline{\textbf{0.926}}&\textbf{0.925}&\textbf{0.925}&\textbf{0.925}&0.923&\underline{\textbf{0.926}}&\textbf{0.925}\\
e+v&0.823 &\underline{\textbf{0.824}}&0.822&0.823&0.822&0.823&0.823&0.821&0.823&0.822\\
endo&0.839 &\textbf{0.840}&0.838&\textbf{0.840}&\textbf{0.840}&\underline{\textbf{0.841}}&0.839&0.838&\underline{\textbf{0.841}}&\underline{\textbf{0.841}}\\
resp&0.723  &\textbf{0.734}&\textbf{0.729}&\textbf{0.737}&\underline{\textbf{0.738}}&\textbf{0.735}& \textbf{0.736}&0.721&\textbf{0.732}&\textbf{0.736}\\
inj&0.607  &\textbf{0.614}&\textbf{0.612}&\underline{\textbf{0.621}}&\textbf{0.616}&\textbf{0.612}&\textbf{0.614}&\textbf{0.608}&\textbf{0.618}&\textbf{0.619}\\
gen&0.659 &\textbf{0.665}&\textbf{0.661}&\underline{\textbf{0.671}}&\textbf{0.669}&\textbf{0.667}&\textbf{0.664}&0.653&\textbf{0.666}&\textbf{0.670}\\
diges&0.653 & \textbf{0.655}&0.648&\textbf{0.655}&\textbf{0.655}&0.651&\textbf{0.655}&0.638&\textbf{0.657}&\underline{\textbf{0.658}}\\
bld&0.489 &\textbf{0.509}&\textbf{0.500}&\textbf{0.507}&\textbf{0.512}&\underline{\textbf{0.513}}&\textbf{0.504}&0.487&\textbf{0.506}&\textbf{0.511}\\
symp&0.413 &\textbf{0.421}&\textbf{0.416}&\textbf{0.414}&\underline{\textbf{0.422}}&\textbf{0.414}&0.403&0.396&\underline{\textbf{0.422}}&\textbf{0.421}\\
ment&0.411 &\textbf{0.427}&\textbf{0.426}&\textbf{0.433}&\textbf{0.442}&\textbf{0.430}&\underline{\textbf{0.447}}&\textbf{0.445}&\textbf{0.434}&\textbf{0.440}\\
nerv&0.442 &\textbf{0.456}&\textbf{0.449}&\textbf{0.454}&\textbf{0.456}&\textbf{0.448}&\textbf{0.446}&0.425&\textbf{0.457}&\underline{\textbf{0.459}}\\
inf&0.587 &\textbf{0.599}&\textbf{0.596}&\textbf{0.597}&\textbf{0.599}&\textbf{0.590}&\textbf{0.598}&\textbf{0.591}&\textbf{0.598}&\underline{\textbf{0.601}}\\
musc&0.139 &0.138&0.120&\textbf{0.144}&\textbf{0.147}&\textbf{0.143}&\underline{\textbf{0.156}}&0.121&\textbf{0.142}&\textbf{0.145}\\
pren&0.578 &\textbf{0.579}&\underline{\textbf{0.580}}&0.578&0.578&0.578&0.578&0.578&\textbf{0.579}&0.578\\
neop&0.610 &\textbf{0.627}&0.610&\textbf{0.623}&\textbf{0.618}&\textbf{0.624}&\textbf{0.623}&\underline{\textbf{0.632}}&\textbf{0.625}&\textbf{0.621}\\
skin&0.163 &\underline{\textbf{0.189}}&0.156&\textbf{0.181}&\textbf{0.174}&\textbf{0.169}&\textbf{0.170}&0.150&\textbf{0.180}&\textbf{0.178}\\
cong&0.152 &\textbf{0.184}&\textbf{0.182}&\textbf{0.196}&\textbf{0.201}&\textbf{0.181}&\underline{\textbf{0.207}}&\textbf{0.170}&\textbf{0.202}&\textbf{0.191}\\
preg&0.234 &\textbf{0.318}&\textbf{0.241}&\textbf{0.312}&\textbf{0.323}&\textbf{0.286}&\textbf{0.306}&0.202&\textbf{0.319}&\underline{\textbf{0.327}}\\
\noalign{\smallskip}\hline
 \end{tabular}}
    \label{tab:ft_par_skip}
\end{table}
\begin{figure}[h]
    \centering
    \includegraphics[width=0.95\textwidth]{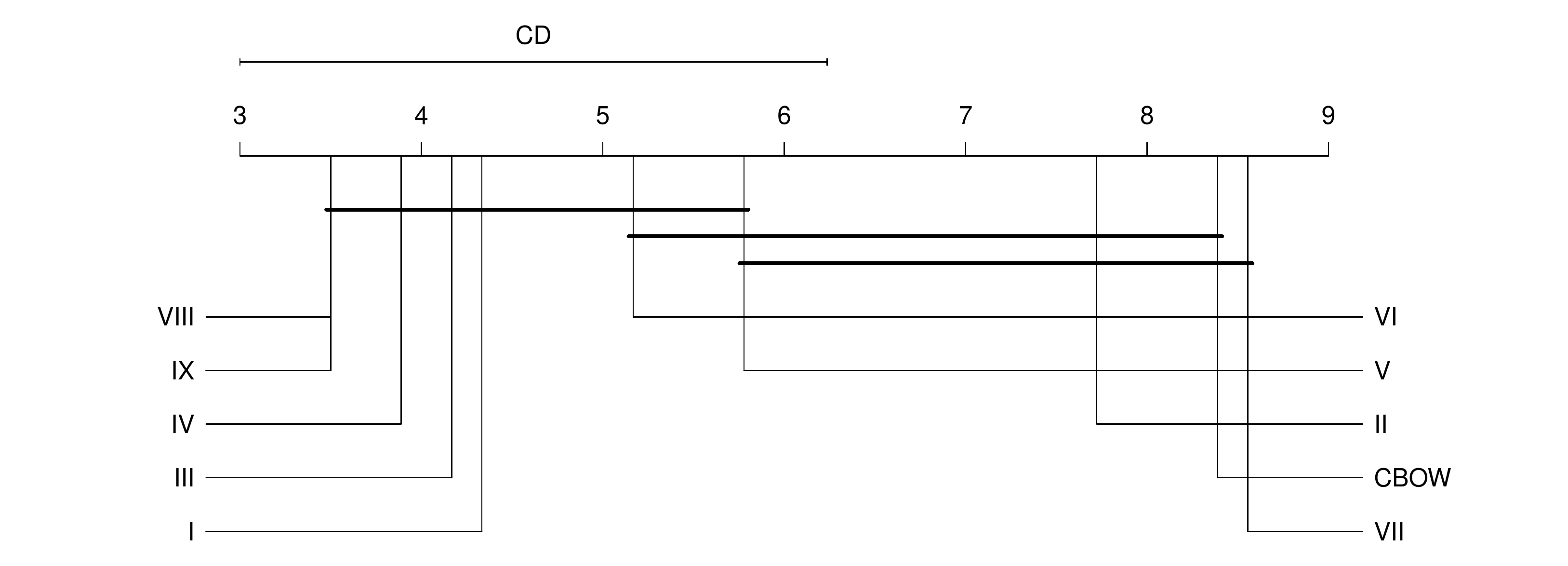}
    \caption{Critical difference plots. Nemenyi post-hoc test (95\% confidence level), identifying statistical differences between learning methods. CBOW values correspond to the best performance for this embedding.}
    \label{fig:cd_plot_skip}
\end{figure}

Table \ref{tab:ft_par_skip} presents a comparison of F-measures of the fastText parameter choices I-IX for both CBOW and Skip-gram embeddings as outlined in Section \ref{sec:fastpar}.  Critical difference plots are available in Fig.~\ref{fig:cd_plot_skip}. Results correspond to 18 embeddings $\times$ 13 classifiers $\times$ 10 folds cv for a total of 2340 tests. The best F-measure among the models using the options presented in Table \ref{table:parameter} for CBOW is also presented. Evidently, for all 18 groups, Skip-gram out-performs CBOW. Option I has the same specifications as the W300 embeddings, and the embeddings presented in Table \ref{tab:models} are trained as per option I for comparison. However, it is evident from Table \ref{tab:ft_par_skip} that varying the fastText parameters impacts F-measures across all 18 ICD-9 groups, but not necessarily always for the better. Thus, care must be taken when selecting these parameters.

\subsubsection{Pre-processing}

\begin{figure}[h]
    \centering
    \includegraphics[width=0.9\textwidth]{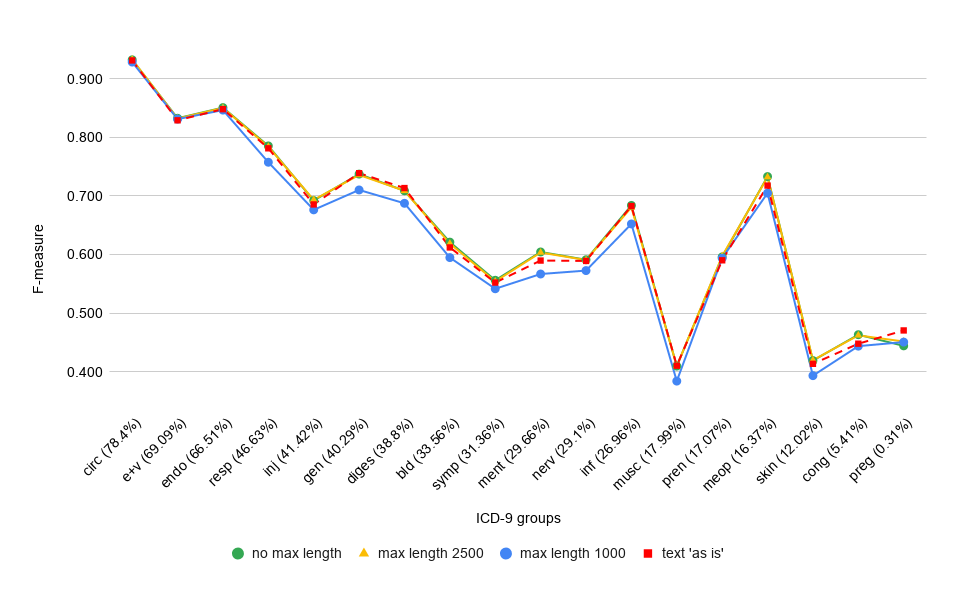}
    \caption{A comparison of F-measures for top level ICD-9 groups between MIMIC III discharge summary used `as is' vs pre-processed. Maximum length indicates pre-processed text being truncated to a maximum token length. Classifier used for experiments is ECC with base classifier of logistic regression, ridge value of one. Embeddings used is T300. Pre-processed text is presented with solid lines and dashes represent text `as is'. }
    \label{fig:prep}
\end{figure}

Figure \ref{fig:prep} presents a comparison of text pre-processed and truncated with the text `as is' within MIMIC III. As mentioned earlier, MIMIC III pre-processes and de-identifies all free-form text released to the public. Here we further process the discharge summary and truncate it to the maximum number of tokens. Generally, the option of discharge summary pre-processed and truncated to 1000 tokens maximum performs much worse than the other options. However, when comparing the text `as is' to the other two pre-processed options, there is very little or no difference in the F-measures. Even for very infrequent categories the differences in F-measure are very marginal. Hence, the question of benefits over trade-off (known and unknown) with regard to pre-processing medical text, or not, remains unclear. It's important to point out that apart from the results presented in this section, all other results presented in this paper are obtained using discharge summaries without any additional pre-processing or truncating other than that already done by MIMIC III.

\subsubsection{Comparison between Embeddings}

\begin{figure}[t]
    \centering
    \includegraphics[width=0.9\textwidth]{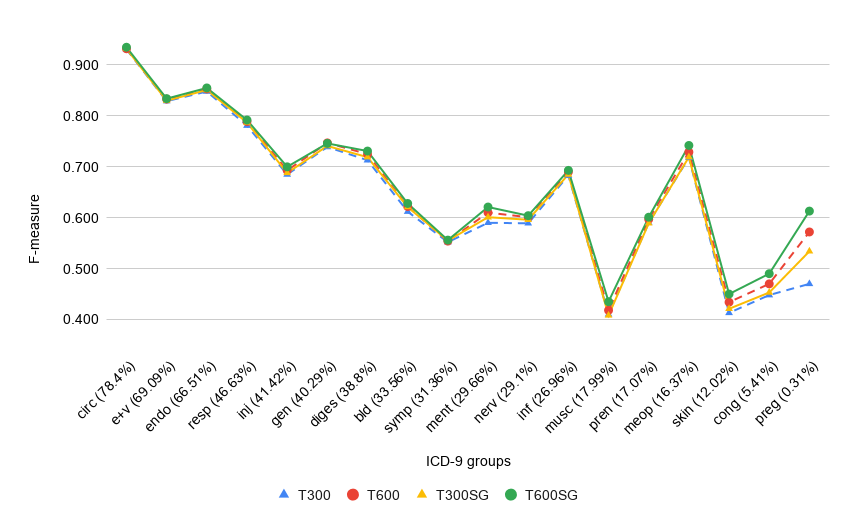}
    \caption{A comparison of F-measures for top level ICD-9 groups presented for embeddings trained using CBOW vs Skip-gram models with dimensions 300 and 600. All experiments used ECC with logistic regression as the base classifier, using a ridge value of one. }
    \label{fig:18comp1}
\end{figure}

\begin{figure}[h!]
    \centering
    \includegraphics[width=0.9\textwidth]{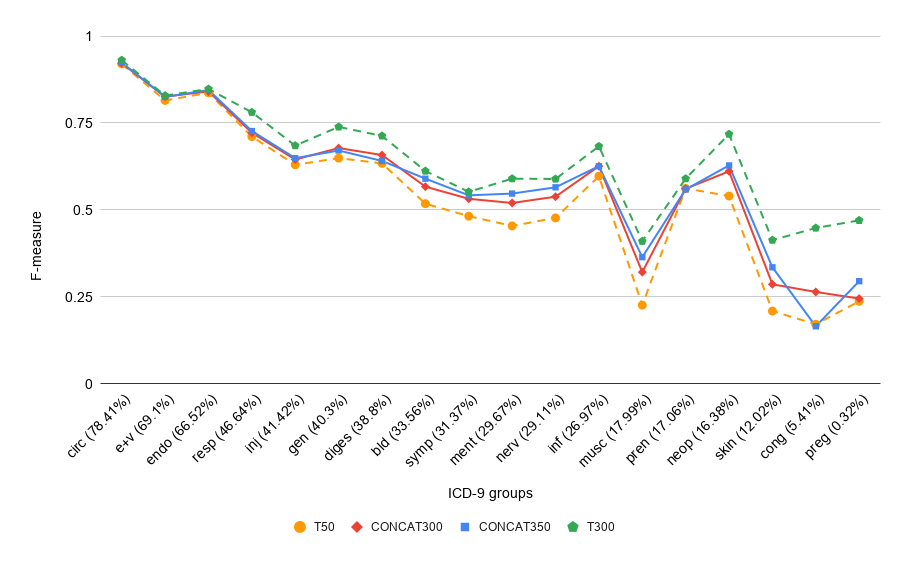}
    \caption{A comparison of F-measures for top level ICD-9 groups is presented. F-measures for two models T50 and T300 are indicated with dashes. Two variations of concatenations CONCAT300 and CONCAT350 are presented with solid lines. All experiments used ECC with logistic regression as the base classifier, using a ridge value of one. }
    \label{fig:split_7}
\end{figure}

Figure \ref{fig:18comp1} presents a comparison of top-level ICD-9 groups between CBOW and Skip-gram models, and between 300 and 600 dimensions. As observed in the binary case, presented in  \cite{yogarajan2020}, increase in dimension does provide an improvement in F-measure. This is more evident with lower frequency groups such as \textsf{skin}, \textsf{cong} and \textsf{preg}. Skip-gram is consistently better than CBOW as observed in Section \ref{sec:fastparresult}. We also compared W300 embeddings, and multi-label variations also present similar observations to that of the binary case, where health-related pre-trained embeddings provide an advantage over general text pre-trained embeddings across all 18 groups.

Figure \ref{fig:split_7} presents a comparison of embeddings formed by concatenating embeddings as per Section \ref{sec:concat}. The base embeddings used here are T50. CONCAT300 is formed by concatenating the embeddings of the statistical outcomes, i.e. CONCAT300 = 50 dim $\times$ (min + max + mean + sd + q1 + q3). CANCAT350 is formed by concatenating the embeddings of the seven text splits 7 $\times$ 50 dim. In comparison, both CONCAT300 and CONCAT350 improve F-measures relative to the base embeddings T50 except for the ICD-9 group \textsf{pren}. CONCAT350 generally performs better than CONCAT300. However, the T300 embeddings outperform both CONCAT300 and CONCAT350 across all 18 groups. Also, the improvements that CONCAT300 and CONCAT350 produce over T50 are not replicated for larger embeddings. For instance when starting with T300 and generating CONCAT1800, or CONCAT2100, no significant improvements are observed. This maybe due to the fact that T300 already performs much better than T50, possibly not leaving much room for further improvement. More future research is needed to investigate this behaviour in more detail.

\subsubsection{Tagging Words}

\begin{figure}[h]
    \centering
    \includegraphics[width=0.9\textwidth]{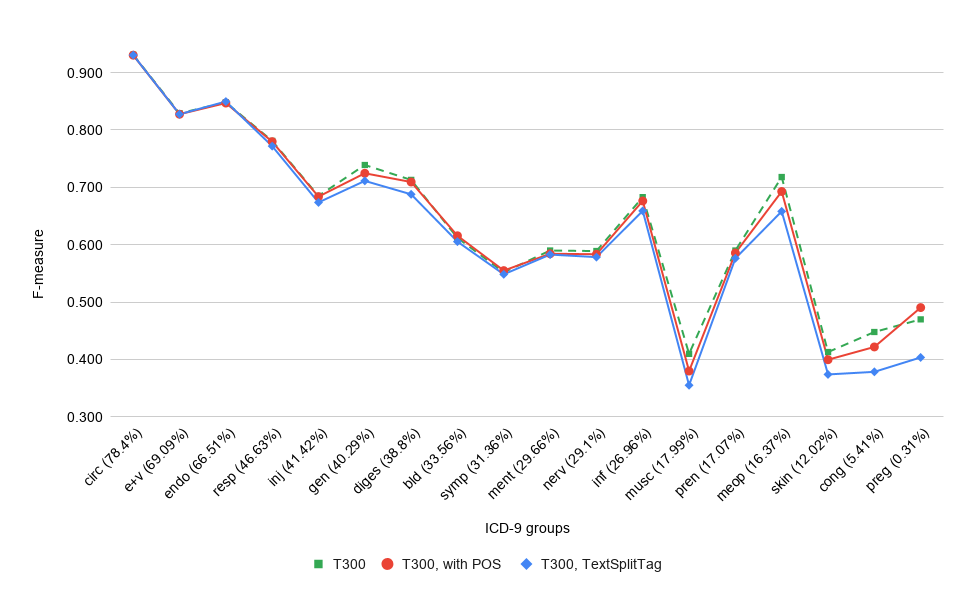}
    \caption{A comparison of F-measures for top level ICD-9 groups between discharge summaries with and without a POS tagger, and with text split tags. All experiments used ECC with logistic regression as the base classifier, using a ridge value of one. }
    \label{fig:pos}
\end{figure}

Figure \ref{fig:pos} presents a comparison of F-measures for top level ICD-9 groups among the MIMIC III discharge summaries with POS tags, with text split tags and for the raw text without any tagging. Evidently, except for categories \textsf{bld}, \textsf{sym}, and  \textsf{preg}, using a POS tagger does not improve the F-measures. For categories \textsf{bld}, \textsf{sym} and \textsf{preg} the use of the POS tagger improves the F-measure, from 0.612 to 0.616, from 0.552 to 0.555, and from 0.470 to 0.491, respectively. When using the text split tagger, the F1 score for \textsf{circ} is equivalent to the no-tagger case, and for category \textsf{endo} there is a small improvement from 0.848 to 0.850.

\subsubsection{Summary: Top level ICD-9 Groups}

\begin{table}[t]
\centering
\caption{A summary of the choices which produce the best F-measure for the top-level ICD-9 groups is presented. }
\resizebox{\linewidth}{!}{    
\begin{tabular}{lllll}\hline\noalign{\smallskip}
ICD-9&ML Classifier&Concatenating Embeddings &Text Tagger&Pre-processing\\
groups & & & &Text \\
\noalign{\smallskip}\hline\noalign{\smallskip}
circ &ECC-SGD,&\footnotesize{T50 $<$ CONCAT300 $<$ CONCAT350 $<$ T300}&no tagger =  &truncated to 2500\\
&  E=100, I = 10 & & TextSplitTag & \\\noalign{\smallskip}
e+v&ECC-SGD,&\footnotesize{T50 $<$ CONCAT350 $<$ CONCAT300 $<$ T300}&no tagger&truncated to 2500, \\
&  E=100, I = 30 & & &no truncate \\\noalign{\smallskip}
endo&ECC-SGD, &\footnotesize{T50 $<$ CONCAT300 $<$ CONCAT350 $<$ T300}&TextSplitTag&no truncate\\
&E=50, I = 100 & & & \\\noalign{\smallskip}
resp&ECC-SGD, &\footnotesize{T50 $<$ CONCAT300 $<$ CONCAT350 $<$ T300}&no tagger&no truncate\\
& E=500, I = 10 & & & \\\noalign{\smallskip}
inj&ECC-LR &\footnotesize{T50 $<$ CONCAT300 $<$ CONCAT350 $<$ T300}&no tagger&truncated to 2500\\
& R = 1, I = 30& & & \\\noalign{\smallskip}
gen&ECC-LR R = 1, &\footnotesize{T50 $<$ CONCAT350 $<$ CONCAT300 $<$ T300}&no tagger&text `as is'\\
& I = 10, 30, 100& & & \\\noalign{\smallskip}
diges&ECC-LR &\footnotesize{T50 $<$ CONCAT350 $<$ CONCAT300 $<$ T300}&no tagger& text `as is'\\
& R = 1, I = 10 & & & \\\noalign{\smallskip}
bld&ECC-LR &\footnotesize{T50 $<$ CONCAT300 $<$ CONCAT350 $<$ T300}&POS&no truncate\\
& R = 1, I = 10 & & & \\\noalign{\smallskip}
symp&ECC-LR &\footnotesize{T50 $<$ CONCAT300 $<$ CONCAT350 $<$ T300}&POS&no truncate\\
& R = 1, I = 10 & & & \\\noalign{\smallskip}
ment&ECC-LR &\footnotesize{T50 $<$ CONCAT300 $<$ CONCAT350 $<$ T300}&no tagger&no truncate\\
& R = 1, I = 10 & & & \\\noalign{\smallskip}
nerv&ECC-LR &\footnotesize{T50 $<$ CONCAT300 $<$ CONCAT350 $<$ T300}&no tagger&no truncate\\
& R = 1, I = 10 & & & \\\noalign{\smallskip}
inf&ECC-LR &\footnotesize{T50 $<$ CONCAT350 = CONCAT300 $<$ T300}&no tagger&no truncate\\
& R = 1, I = 10 & & & \\\noalign{\smallskip}
musc&ECC-LR &\footnotesize{T50 $<$ CONCAT300 $<$ CONCAT350 $<$ T300}&no tagger&no truncate\\
& R = 1, I = 10 & & & \\\noalign{\smallskip}
pren&BR, Log, &\footnotesize{CONCAT350 $<$ CONCAT300 $<$ T50 $<$ T300}&no tagger&truncated to 2500\\
& R = 1 & & & \\\noalign{\smallskip}
neop&ECC-LR &\footnotesize{T50 $<$ CONCAT300 $<$ CONCAT350 $<$ T300}&no tagger&no truncate\\
& R = 1, I = 10 & & & \\\noalign{\smallskip}
skin&ECC-LR &\footnotesize{T50 $<$ CONCAT300 $<$ CONCAT350 $<$ T300}&no tagger&truncated to 2500\\
& R = 1, I = 10 & & & \\\noalign{\smallskip}
cong&ECC-LR &\footnotesize{T50 $<$ CONCAT350 $<$ CONCAT300 $<$ T300}&no tagger&no truncate\\
& R = 1, I = 30 & & & \\\noalign{\smallskip}
preg&BR-LR, &\footnotesize{T50 $<$ CONCAT300 $<$ CONCAT350 $<$ T300}&POS&text `as is'\\
& R = 1 & & & \\
\noalign{\smallskip}\hline
\end{tabular}}
\label{tab:summary}
\end{table}

The 18 top-level ICD-9 groups are used to present experimental results of several variations of techniques as outlined in Section \ref{sec:methods}. Skip-gram models improve the F-measures of all 18 groups compared to CBOW. Also, we presented comparisons of several variations of parameters to pre-trained embeddings for both CBOW and Skip-gram. Results to such modifications indicate changes to F-measures across all 18 labels, but not always for the better (see section \ref{sec:fastparresult} for details of results). The T600SG embeddings are the best-performing choice for all but one of the groups. It is only for category \textsf{gen} that the CBOW-based T600 manages to outperform T600SG. As observed in the binary case, embeddings trained using health-related data do provide an advantage over general text pre-trained embeddings. Also, higher dimensions improve F-measures, especially evident in low-frequency categories. A summary of the choices which provided the best F-measure for each ICD-9 top-level groups is presented in Table \ref{tab:summary}.

\begin{figure}[h!]
    \centering
    \includegraphics[width=0.9\textwidth]{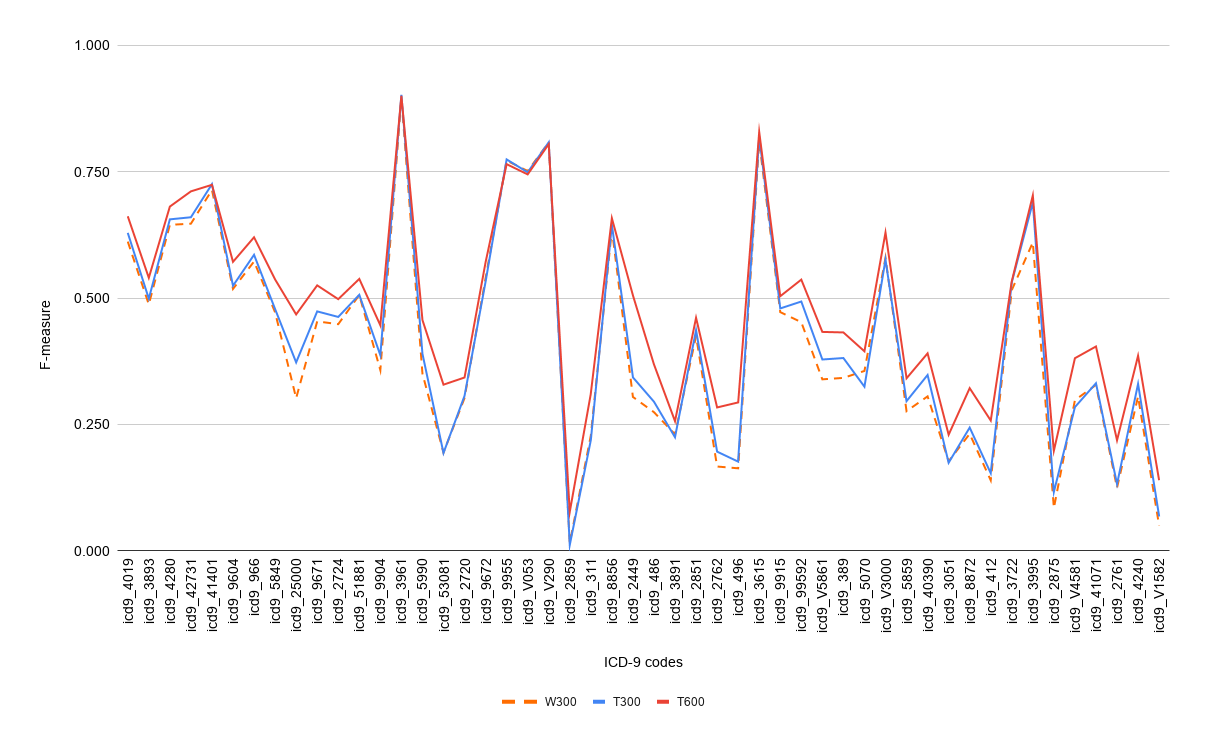}
    \caption{A comparison of F-measures for the top 50 most frequently occurring ICD-9 codes ordered from the highest frequency ICD-9 code 401.9 down to the 50th V15.82, is presented. Comparisons are between the W300 and T300 embeddings as well as the T300 and T600 embeddings. All experiments used ECC with logistic regression as the base classifier, using a ridge value of one. }
    \label{fig:top50dim_gen}
\end{figure}

\begin{figure}[h!]
    \centering
    \includegraphics[width=0.9\textwidth]{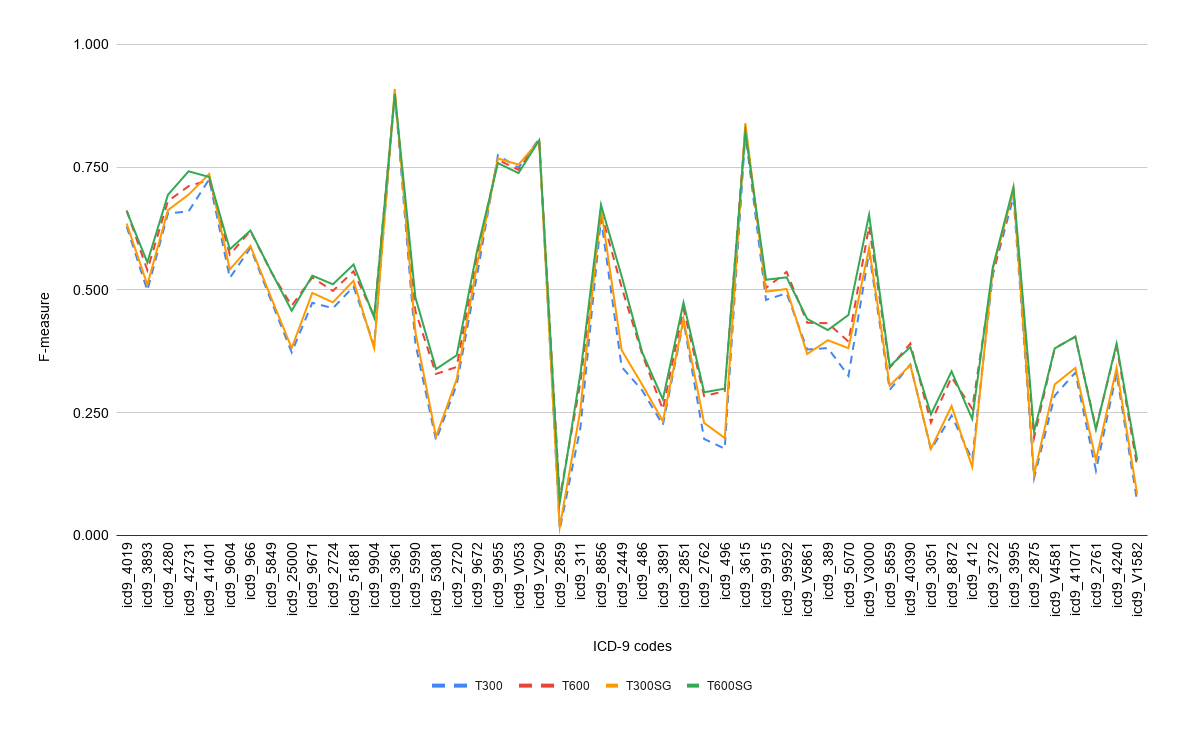}
    \caption{A comparison of F-measures for most frequently occurring top 50 ICD-9 codes ordered from the highest frequent ICD-9 code 401.9 to the 50th V15.82 is presented. Comparisons are between CBOW trained embeddings T300 and T600 (solid lines) with Skip-gram trained embeddings T300SG and T600SG (dashes).  All experiments used ECC with logistic regression as the base classifier, using a ridge value of one.}
    \label{fig:top50_cbowskip}
\end{figure}

\begin{figure}[h!]
    \centering
    \includegraphics[width=0.9\textwidth]{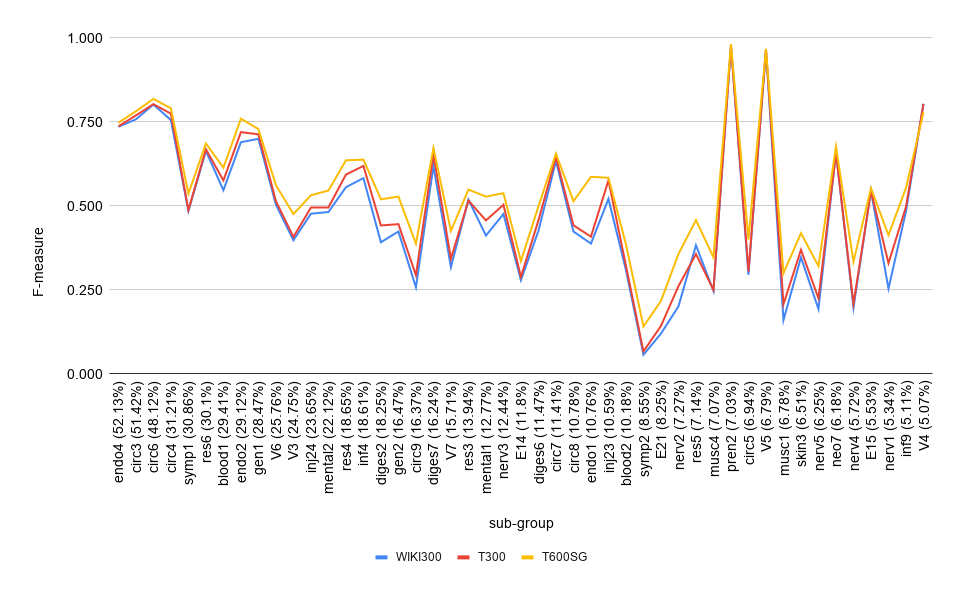}
    \caption{A comparison of F-measures between W300 and T300, and between 300 and 600 dimensions is presented for ICD-9 sub-level groups occurring in more than 5\% of the cases in  MIMIC III.  All experiments used ECC with logistic regression as the base classifier, using a ridge value of one.}
    \label{fig:sub1}
\end{figure}

\begin{figure}[h!]
    \centering
    \includegraphics[width=0.9\textwidth]{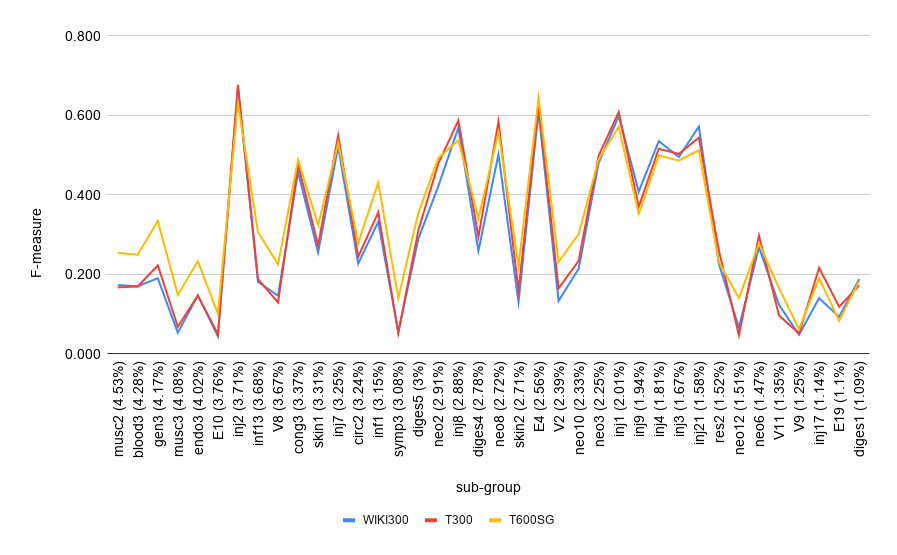}
  \caption{A comparison of F-measures between W300 and T300, and between 300 and 600 dimensions is presented for ICD-9 sub-level groups occurring in between 1\% and 5\% of the cases in MIMIC III. All experiments used ECC with logistic regression as the base classifier, using a ridge value of one.}
    \label{fig:sub2}
\end{figure}

\begin{figure}[h!]
    \centering
    \includegraphics[width=0.9\textwidth]{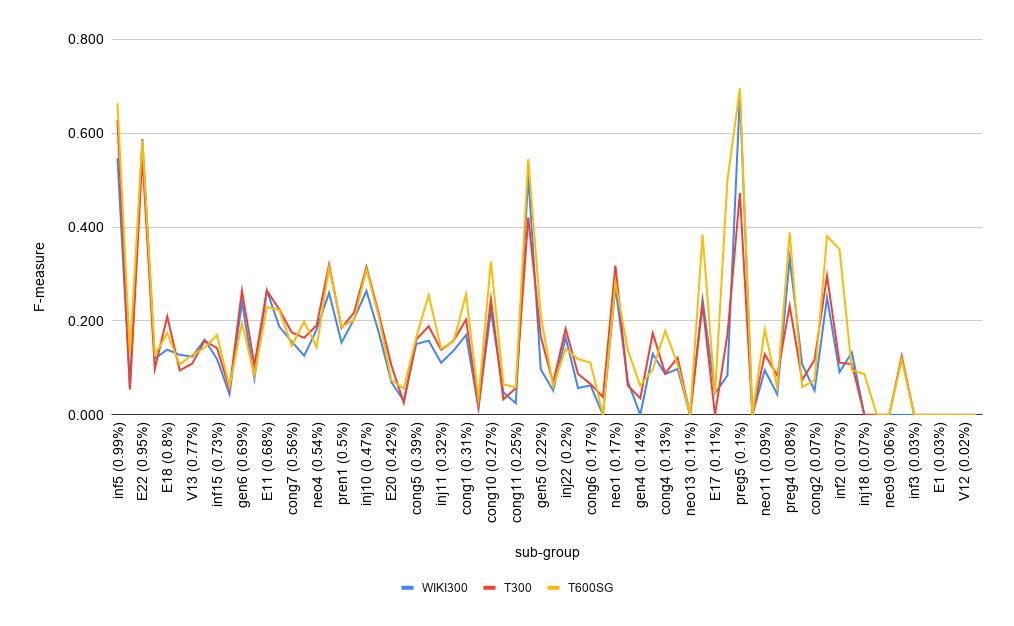}
  \caption{A comparison of F-measures between W300 and T300, and between 300 and 600 dimensions is presented for ICD-9 sub-level groups occuring in less than 1\% of the cases  in MIMIC III.  All experiments used ECC with logistic regression as the base classifier, using a ridge value of one.}
    \label{fig:sub3}
\end{figure}

\subsection{Highest Frequency Medical Codes}\label{sec:top50}

This section presents the results for the top 50 most frequent individual ICD-9 codes. Mullenbach et al. (2018) \cite{mullenbach2018explainable}, arguably the most prominent work in predicting ICD-9 codes from MIMIC III, considers the top 50 most frequent codes from both diagnosis and procedure ICD-9 codes. Hence, we also present results for the same top 50 ICD-9 codes where the most frequent code is 401.9 and occurs in 35.13\% of all cases, whereas the least frequent code V15.82 is only present in about 5\% of all cases (see Table \ref{table:summary_stats} for more details). 

Figure \ref{fig:top50dim_gen} presents a comparison of F-measures between general text pre-trained embeddings W300 and health-related pre-trained T300 and T600 embeddings for the 50 topmost frequently occurring ICD-9 codes. As observed with the 18 top-level ICD-9 groups, when compared to general text trained embeddings, F-measures of health-related trained embeddings are better. Also an increase in dimensions from 300 to 600 results in a considerable improvement in F-measure, far more than what was noticed in the 18 label case. For example, for ICD-9 code 530.81 the F-measure improves from 0.194 to 0.329, and for 410.71 from 0.332 to 0.404.       

Figure \ref{fig:top50_cbowskip} presents a comparison between CBOW and Skip-gram trained embeddings for the 50 topmost frequently occurring ICD-9 codes. As with the 18 label case, in general Skip-gram models are better than CBOW models, except for a few ICD-9 codes, such as 995.5, or 389.

\subsection{Sub-level Groups of Medical Codes}

A comparison of F-measures for sub-level ICD-9 groups is presented where the 155 labels are treated as one multi-label classification problem. We only use the sub-groups which are recorded in more than ten unique hospital admissions, hence 155 and not the entire 167 possible sub-groups. F-measures are presented from highest frequency of occurrence to the lowest. Figure \ref{fig:sub1} presents results for ICD-9 sub-groups with occurrences of more than 5\%, Figure \ref{fig:sub2} for sub-level groups with occurrences between 1\% and 5\%, and Figure \ref{fig:sub3} contains sub-groups with less than 1\% occurrences. A comparison of F-measures between W300 and T300, and between 300 and 600 dimensions is presented.

For most of the 155 sub-groups T300 outperforms W300, and for most sub-level groups, there is a definite improvement in F-measure when increasing the number of dimensions from 300 up to 600. This pattern matches the results for 18 and for 50 labels as presented above in sections \ref{sec:top} and \ref{sec:top50}.

\subsection{Overall Results}

In this section we present results for the overall performance of a multi-label medical text classification problem with 18, 50 and 155 labels. We present micro-averaged and macro-average F1 measures, aligned with prior work (see section \ref{sec:lit} for examples).

Table \ref{tab:micromacro} presents micro- and macro-averaged F-measures for 18, 50 and 155 label multi-label medical text classification tasks with embeddings variations. The overall pattern matches the observations for individual label-level F-measures, where the combination of higher dimensions and a Skip-gram model usually results in the highest performance measures. For all three groups of labels (18, 50, or 155), micro- and macro-averaged F1 scores are always better for T300 than for W300.    

\begin{table}[t]
\centering
\caption{Micro- and macro- averaged F-measures for multi-label medical text classification problem with 18, 50 and 155 labels is presented. All experiments used ECC with logistic regression as the base classifier, using a ridge value of one, except for ** where the classifier is explicitly stated. * refers to ``text pre-processed and truncated to 2500 tokens'' as per section \ref{sec:preprocess}. Bold is used to indicate the best measures for each case. }
\resizebox{\linewidth}{!}{    
\begin{tabular}{lllllll}
\hline\noalign{\smallskip}
Model Description & Micro F1 & Macro F1 & \qquad & Model & Micro F1 & Macro F1 \\
\noalign{\smallskip}\hline\noalign{\smallskip}
\multicolumn{3}{c}{\textit{Top-level: 18 labels}} & & \multicolumn{3}{c}{\textit{Top 50: 50 labels}} \\
W300&0.730&0.648  & & W300&0.484&0.434\\
T300&0.734&0.653  & & T300&0.497&0.445 \\
T300SG& 0.737& 0.658 & & T600&0.532&0.486\\
T300, truncated to 2500* &0.737 &0.654 & & T600SG&\textbf{0.539}&\textbf{0.493}\\
CONCAT300 &0.676 & 0.589 & & & & \\
CONCAT350 &0.702 & 0.593 & & \multicolumn{3}{c}{\textit{Sub-level: 155 labels}}\\
T300, POS tag &0.730 &0.646 & & W300&0.534&0.293\\
T300, TextSplitTag &0.723 & 0.684 & &  T300&0.551&0.306\\
T300, ECC-SGD, E=500, I=100** & 0.721 & 0.634 & & T600SG&\textbf{0.568}&\textbf{0.337}\\
T600&0.742&0.665 &  & & &\\
T600SG&\textbf{0.745}&\textbf{0.674}& & & & \\
\noalign{\smallskip}\hline
\end{tabular}}
\label{tab:micromacro}
\end{table}

\section{Conclusions}

We present a detailed analysis of clinical NLP techniques used to enhance the embeddings layer of a multi-label medical text classification task. We focus on predicting ICD-9 for patients with multi-morbidity, and present results for 18, 50 and 155 labels. Results and analysis are primarily done at individual label level. Given the imbalanced nature of the data, at the individual label level, it is evident that variations in embeddings such as the use of Skip-gram model over CBOW, and higher dimensional embeddings do result in improvements in F-measure. These improvements are more significant with less frequent labels. This is evident across all three setups, regardless of using 18, 50 or 155 labels. These improvements and differences are also evident in overall micro- and macro-averaged F-measures. This research emphasises the need for enhancing text representations, and results show that there is a definite benefit in incorporating additional features. The benefits are depended on the data distributions and the task at hand. This paper used predicting medical codes as an example. However, the NLP techniques used in this research can be adapted to other tasks where multi-label medical text classification is required.     

Our analysis on pre-processed text only presents marginal improvements to F-measure when compared to text `as is'. Hence there is no clear indication that additional pre-processing is required on already pre-processed and de-identified data, such as MIMIC III, especially given the nature of the medical text.

\bibliographystyle{spmpsci}  
\bibliography{references}

\begin{thebibliography}{10}
\providecommand{\url}[1]{{#1}}
\providecommand{\urlprefix}{URL }
\expandafter\ifx\csname urlstyle\endcsname\relax
  \providecommand{\doi}[1]{DOI~\discretionary{}{}{}#1}\else
  \providecommand{\doi}{DOI~\discretionary{}{}{}\begingroup
  \urlstyle{rm}\Url}\fi

\bibitem{abacha2019overview}
Abacha, A.B., Shivade, C., Demner-Fushman, D.: {Overview of the MEDIQA 2019
  shared task on textual inference, question entailment and question
  answering}.
\newblock In: {Proceedings of the 18th BioNLP Workshop and Shared Task}, pp.
  370--379 (2019)

\bibitem{aubert2019patterns}
Aubert, C.E., Schnipper, J.L., Fankhauser, N., Marques-Vidal, P., Stirnemann,
  J., Auerbach, A.D., Zimlichman, E., Kripalani, S., Vasilevskis, E.E.,
  Robinson, E., et~al.: Patterns of multimorbidity associated with 30-day
  readmission: a multinational study.
\newblock BMC public health \textbf{19}(1), 738 (2019)

\bibitem{baumel2018multi}
Baumel, T., Nassour-Kassis, J., Cohen, R., Elhadad, M., Elhadad, N.:
  Multi-label classification of patient notes: case study on icd code
  assignment.
\newblock In: Workshops at the Thirty-Second AAAI Conference on Artificial
  Intelligence (2018)

\bibitem{NLTK09}
Bird, S., Klein, E., Loper, E.: {Natural Language Processing with Python}.
\newblock O'Reilly Media (2009)

\bibitem{bojanowski2016enriching}
Bojanowski, P., Grave, E., Joulin, A., Mikolov, T.: Enriching word vectors with
  subword information.
\newblock arXiv preprint arXiv:1607.04606  (2016)

\bibitem{data2016secondary}
Data, M.C.: Secondary Analysis of Electronic Health Records.
\newblock Springer (2016)

\bibitem{Demsar2006}
Dem\v{s}ar, J.: Statistical comparisons of classifiers over multiple data sets.
\newblock Journal of Machine Learning Research \textbf{7}, 1--30 (2006)

\bibitem{du2019ml}
Du, J., Chen, Q., Peng, Y., Xiang, Y., Tao, C., Lu, Z.: Ml-net: multi-label
  classification of biomedical texts with deep neural networks.
\newblock Journal of the American Medical Informatics Association
  \textbf{26}(11), 1279--1285 (2019)

\bibitem{flegel2018we}
Flegel, K.: What we need to learn about multimorbidity.
\newblock CMAJ \textbf{190}(34) (2018)

\bibitem{Garcia2009}
Garcia, S., Herrera, F.: An extension on "statistical comparisons of
  classifiers over multiple data sets" for all pairwise comparisons.
\newblock Journal of Machine Learning Research \textbf{9}, 2677--2694 (2009)

\bibitem{garla2011yale}
Garla, V., Re~III, V.L., Dorey-Stein, Z., Kidwai, F., Scotch, M., Womack, J.,
  Justice, A., Brandt, C.: {The Yale cTAKES extensions for document
  classification: architecture and application}.
\newblock Journal of the American Medical Informatics Association
  \textbf{18}(5), 614--620 (2011)

\bibitem{godbole2004discriminative}
Godbole, S., Sarawagi, S.: Discriminative methods for multi-labeled
  classification.
\newblock In: Pacific-Asia conference on knowledge discovery and data mining,
  pp. 22--30. Springer (2004)

\bibitem{goldberg2017neural}
Goldberg, Y.: Neural network methods for natural language processing.
\newblock Synthesis Lectures on Human Language Technologies \textbf{10}(1),
  1--309 (2017)

\bibitem{goldberger2000physiobank}
Goldberger, A.L., Amaral, L.A., Glass, L., Hausdorff, J.M., Ivanov, P.C., Mark,
  R.G., Mietus, J.E., Moody, G.B., Peng, C.K., Stanley, H.E.: {PhysioBank,
  PhysioToolkit, and PhysioNet: components of a new research resource for
  complex physiologic signals}.
\newblock Circulation \textbf{101}(23), e215--e220 (2000)

\bibitem{grave2018learning}
Grave, E., Bojanowski, P., Gupta, P., Joulin, A., Mikolov, T.: Learning word
  vectors for 157 languages.
\newblock In: Proceedings of the International Conference on Language Resources
  and Evaluation (LREC 2018) (2018)

\bibitem{hausmann2019sensitivity}
Hausmann, D., Kiesel, V., Zimmerli, L., Schlatter, N., von Gunten, A.,
  Wattinger, N., Rosemann, T.: Sensitivity for multimorbidity: The role of
  diagnostic uncertainty of physicians when evaluating multimorbid video
  case-based vignettes.
\newblock PloS one \textbf{14}(4) (2019)

\bibitem{huggard2019feature}
Huggard, H., Zhang, A., Zhang, E., Koh, Y.S.: Feature importance for biomedical
  named entity recognition.
\newblock In: Australasian Joint Conference on Artificial Intelligence, pp.
  406--417. Springer (2019)

\bibitem{jensen2012}
Jensen, P.B., Jensen, L.J., Brunak, S.: Mining electronic health records:
  towards better research applications and clinical care.
\newblock Nature Reviews Genetics \textbf{13}(6), 395 (2012)

\bibitem{johnson2017reproducibility}
Johnson, A.E., Pollard, T.J., Mark, R.G.: Reproducibility in critical care: a
  mortality prediction case study.
\newblock In: Machine Learning for Healthcare Conference, pp. 361--376 (2017)

\bibitem{johnson2016mimic}
Johnson, A.E., Pollard, T.J., Shen, L., Li-wei, H.L., Feng, M., Ghassemi, M.,
  Moody, B., Szolovits, P., Celi, L.A., Mark, R.G.: {MIMIC-III, a freely
  accessible critical care database}.
\newblock Scientific data \textbf{3}, 160035 (2016)

\bibitem{joulin2016fasttext}
Joulin, A., Grave, E., Bojanowski, P., Douze, M., J{\'e}gou, H., Mikolov, T.:
  Fasttext.zip: Compressing text classification models.
\newblock arXiv preprint arXiv:1612.03651  (2016)

\bibitem{joulin2016bag}
Joulin, A., Grave, E., Bojanowski, P., Mikolov, T.: Bag of tricks for efficient
  text classification.
\newblock arXiv preprint arXiv:1607.01759  (2016)

\bibitem{lee2019ncuee}
Lee, L.H., Lu, Y., Chen, P.H., Lee, P.L., Shyu, K.K.: {NCUEE at MEDIQA 2019:
  Medical Text Inference Using Ensemble BERT-BiLSTM-Attention Model}.
\newblock In: Proceedings of the 18th BioNLP Workshop and Shared Task, pp.
  528--532 (2019)

\bibitem{li2018automated}
Li, M., Fei, Z., Zeng, M., Wu, F., Li, Y., Pan, Y., Wang, J.: Automated {ICD-9}
  coding via a deep learning approach.
\newblock IEEE/ACM transactions on computational biology and bioinformatics
  (2018)

\bibitem{liu2013integrated}
Liu, H., Wagholikar, K.B., Jonnalagadda, S., Sohn, S.: {Integrated cTAKES for
  Concept Mention Detection and Normalization.}
\newblock In: CLEF (Working Notes) (2013)

\bibitem{DBLP:journals/corr/abs-1301-3781}
Mikolov, T., Chen, K., Corrado, G., Dean, J.: Efficient estimation of word
  representations in vector space.
\newblock CoRR \textbf{abs/1301.3781} (2013).
\newblock \urlprefix\url{http://arxiv.org/abs/1301.3781}

\bibitem{mori2019associations}
Mori, T., Hamada, S., Yoshie, S., Jeon, B., Jin, X., Takahashi, H., Iijima, K.,
  Ishizaki, T., Tamiya, N.: The associations of multimorbidity with the sum of
  annual medical and long-term care expenditures in japan.
\newblock BMC geriatrics \textbf{19}(1), 69 (2019)

\bibitem{mullenbach2018explainable}
Mullenbach, J., Wiegreffe, S., Duke, J., Sun, J., Eisenstein, J.: Explainable
  prediction of medical codes from clinical text.
\newblock arXiv preprint arXiv:1802.05695  (2018)

\bibitem{purushotham2017benchmark}
Purushotham, S., Meng, C., Che, Z., Liu, Y.: Benchmark of deep learning models
  on large healthcare mimic datasets.
\newblock arXiv preprint arXiv:1710.08531  (2017)

\bibitem{JesseRead2009}
Read, J., Pfahringer, B., Holmes, G., Frank, E.: Classifier chains for
  multi-label classification.
\newblock In: Joint European Conference on Machine Learning and Knowledge
  Discovery in Databases, pp. 254--269. Springer (2009)

\bibitem{JesseRead2011}
Read, J., Pfahringer, B., Holmes, G., Frank, E.: Classifier chains for
  multi-label classification.
\newblock Machine learning \textbf{85}(3), 333 (2011)

\bibitem{read2016}
Read, J., Reutemann, P., Pfahringer, B., Holmes, G.: {MEKA: A
  Multi-label/Multi-target Extension to WEKA}.
\newblock {Journal of Machine Learning Research} \textbf{17}(21), 1--5 (2016).
\newblock \urlprefix\url{http://jmlr.org/papers/v17/12-164.html}

\bibitem{reategui2018comparison}
Re{\'a}tegui, R., Ratt{\'e}, S.: {Comparison of MetaMap and cTAKES for entity
  extraction in clinical notes}.
\newblock {BMC Medical Informatics and Decision Making} \textbf{18}(3), 74
  (2018)

\bibitem{rios2018emr}
Rios, A., Kavuluru, R.: {EMR} coding with semi--parametric multi--head matching
  networks.
\newblock In: Proceedings of the conference. Association for Computational
  Linguistics. North American Chapter. Meeting, vol. 2018, p. 2081. NIH Public
  Access (2018)

\bibitem{roberts2017overview}
Roberts, K., Demner-Fushman, D., Voorhees, E.M., Hersh, W.R., Bedrick, S.,
  Lazar, A.J., Pant, S.: Overview of the {TREC} 2017 precision medicine track.
\newblock NIST Special Publication pp. 500--324 (2017)

\bibitem{ryan2018beyond}
Ryan, B.L., Jenkyn, K.B., Shariff, S.Z., Allen, B., Glazier, R.H., Zwarenstein,
  M., Fortin, M., Stewart, M.: Beyond the grey tsunami: a cross-sectional
  population-based study of multimorbidity in ontario.
\newblock Canadian Journal of Public Health \textbf{109}(5-6), 845--854 (2018)

\bibitem{savova2010mayo}
Savova, G.K., Masanz, J.J., Ogren, P.V., Zheng, J., Sohn, S., Kipper-Schuler,
  K.C., Chute, C.G.: {Mayo clinical Text Analysis and Knowledge Extraction
  System (cTAKES): architecture, component evaluation and applications}.
\newblock Journal of the American Medical Informatics Association
  \textbf{17}(5), 507--513 (2010)

\bibitem{tsoumakas2007multi}
Tsoumakas, G., Katakis, I.: Multi-label classification: An overview.
\newblock International Journal of Data Warehousing and Mining (IJDWM)
  \textbf{3}(3), 1--13 (2007)

\bibitem{yang2017uevora}
Yang, H., Gon{\c{c}}alves, T.: {UEvora at CLEF eHealth 2017 Task 3.}
\newblock In: CLEF (Working Notes) (2017)

\bibitem{yogarajan2020}
Yogarajan, V., Gouk, H., Smith, T., Mayo, M., Pfahringer, B.: {Comparing High
  Dimensional Word Embeddings Trained on Medical Text to Bag-of-Words For
  Predicting Medical Codes.}
\newblock {Proceedings of the Asian Conference on Intelligent Information and
  Database Systems (ACIIDS 2020). In N. T. Nguyen et al. (Eds.), Lecture Notes
  on Artificial Intelligence (LNAI), Springer Nature. (to appear).}
  \textbf{12033}, 1--12 (2020)

\bibitem{yogarajan2020review}
Yogarajan, V., Pfahringer, B., Mayo, M.: A review of automatic end-to-end
  de-identification: Is high accuracy the only metric?
\newblock Applied Artificial Intelligence pp. 1--19 (2020)

\bibitem{zeng2019automatic}
Zeng, M., Li, M., Fei, Z., Yu, Y., Pan, Y., Wang, J.: Automatic icd-9 coding
  via deep transfer learning.
\newblock Neurocomputing \textbf{324}, 43--50 (2019)

\bibitem{zhang2005k}
Zhang, M.L., Zhou, Z.H.: {A k-nearest neighbor based algorithm for multi-label
  classification}.
\newblock In: 2005 IEEE international conference on granular computing, vol.~2,
  pp. 718--721. IEEE (2005)

\end{thebibliography}

\end{document}